\documentclass[journal]{IEEEtran}
\input{decl}

\begin{document}

\title{Bandlimited signal reconstruction\\
from orthogonally-projected data}

\author{Nguyen T. Thao,~\IEEEmembership{Member,~IEEE,} and Marek Mi\'skowicz,~\IEEEmembership{Senior Member,~IEEE} 
\thanks{N. T. Thao is with the Department of Electrical Engineering, The City College of New York, CUNY, New York, USA, email: tnguyen@ccny.cuny.edu.}
\thanks{M. Mi\'skowicz is with the Department of Measurement and Electronics, AGH University of Krak\'ow, Krak\'ow, Poland, email: miskow@agh.edu.pl}
\thanks{M. Mi\'skowicz was supported by the Polish National Center of Science under grants DEC-2018/31/B/ST7/03874.}
}

\maketitle


\begin{abstract}
We show that a broad class of signal acquisition schemes can be interpreted as recording data from a signal $x$ in a space $\scU$ (typically, though not exclusively, a space of bandlimited functions) via an orthogonal projection $w = P_{\scV} x$ onto another space $\scV$. A basic reconstruction method in this case consists in alternating projections between the input space $\scU$ and the affine space $ \scW$ of signals $u$ satisfying $P_{\scV} u = w$ (POCS method).
Although this method is classically known to be slow, our work reveals new insights and contributions:
(i) it applies to new complex encoders emerging from event-based sampling, for which no faster reconstruction method is currently available;
(ii) beyond perfect reconstruction, it converges robustly under insufficient (e.g., sub-Nyquist) or inconsistent data (due to noise or errors);
(iii) the limit of convergence achieves optimal least-squares approximations under such conditions;
(iv) semi-convergence inherently results in regularized reconstructions under ill-posed data acquisition conditions; (v) when $w$ is produced by discrete sampling, the iterative method can be rigorously discretized for DSP implementation — even with non-separable input spaces $\scU$.
While moving beyond the traditional focus on perfect reconstruction in harmonic analysis, our analysis preserves the deterministic framework of infinite-dimensional Hilbert spaces, consistent with Shannon's sampling theory. For illustration, we apply our proposed theory to two contrasting sampling situations: multi-channel time encoding and nonuniform point sampling in a Sobolev space.
\end{abstract}

\section{Introduction}\label{sec:intro}

The growing interest in event-based sampling \cite{Miskowicz2018} has renewed attention to signal reconstruction from
nonuniform samples. By tradition, the theoretical research on this subject had been mostly focused on perfect reconstruction of bandlimited signals from point samples \cite{Duffin52,Benedetto92,Feichtinger94,Marvasti01,benedetto2001modern}. However, in an effort to improve the resolution of data acquisition circuits, event-based sampling has expanded into developing encoders that acquire generalized forms of samples, for which little theoretical result of signal recovery had been published (with the exception of \cite{Grochenig92b}). Such a direction experienced significant growth with the pioneering work of Lazar and Tóth in \cite{Lazar04}, who demonstrated the possibility of perfect signal reconstruction from samples that consist of input integrals over successive but nonuniform intervals. This type of sampling is realized by asynchronous Sigma-Delta modulation (ASDM) or by integrate-and-fire encoding \cite{lazar2008faithful,florescu2015novel,Adam20}. The proposed reconstruction algorithm, which we will call the LT method, was an adaptation of an iterative algorithm originally designed in \cite[\S8.4]{Feichtinger94} for point sampling. This growing research is however limited in two aspects: (a) a difficulty to generalize the LT algorithm for sampling schemes beyond the pure integration of a bandlimited signal; (b) a lack of theoretical framework for the analysis of the LT method under noisy or insufficient sampling. On the first aspect, numerical results from \cite{Thao22a} showed that the LT algorithm is no longer guaranteed to converge under leaky integration even with oversampling. A proposed extension of this algorithm to multi-channel signal encoding in \cite{Adam20} did yield numerical convergence, but as an empirical observation without existing theoretical justifications. On the second aspect, the LT algorithm remains the result of research in harmonic analysis specifically geared to situations of perfect reconstruction, as mentioned above. The analysis of noisy and insufficient sampling is however of major importance in practice. Note the basic fact that the finite sampling of a bandlimited signal is automatically insufficient in the traditional $L^2(\RR)$ space of Shannon's sampling theory.

In this article, we propose a theoretical framework of sampling schemes that have potential applications in event-based sampling and benefit from an available iterative algorithm of bandlimited input reconstruction, with the following property:  it systematically converges to the best input approximation from the available samples whenever the sampling problem is theoretically well-posed. This implies {\em a fortiori} that the algorithm achieves perfect reconstruction whenever the sampling data is exact and uniquely characteristic of the input. Meanwhile, within the same theoretical framework, it addresses the practical concerns of noisy and insufficient sampling.

Our framework is originated from \cite{Thao21a} where an alternative iterative algorithm to the LT method was proposed for bandlimited input reconstruction from integrals. While the algorithm yielded similar complexity and rate of convergence to the LT method in the situation where the latter achieves perfect reconstruction, the convergence to optimal input approximations was obtained under any other stable sampling situations. The algorithm was based on the method of projections onto convex sets (POCS) \cite{Youla82,Combettes93,Bauschke96}. The task of this article is to identify the most fundamental principles behind the method of \cite{Thao21a} to build up the proposed general framework, and show initial applications of this generalization.

Assume that the input $x$ to be acquired is bandlimited, or more generally belongs to some closed space $\scU$ of signals. It is observed in this article that the method of \cite{Thao21a} with its outstanding convergence properties is applicable to any encoding scheme that acquires data of the type
\begin{equation}\label{data-proj}
w:=P_\scV\' x,
\end{equation}
where $\scV$ is a another space of signals within a given ambient space $\scH$, and $P_\scV$ is the orthogonal projection onto $\scV$. The associated reconstruction algorithm consists in alternating orthogonal projections between $\scU$ and the space of $\scW$ of all signals $u$ such that $P_\scV u=w$, to form the sequence of estimates
\begin{equation}\label{POCS-abs}
x\up{n+1}=P_\scU P_\scW\'x\up{n},\qquad n\geq0.
\end{equation}
This iteration is classically known to converge to an element of $\scU\cap\scW$, which contains $x$ by construction. Perfect reconstruction is thus automatically achieved when $\scU\cap\scW$ is a singleton. One actually recognizes the well-known Papoulis-Gerchberg algorithm \cite{Papoulis75,Gerchberg74} for bandlimited extrapolation in the case where $\scU$ is a space of bandlimited signals and $\scV$ is a space of time-limited signals.
In this paper, however, we review in details the properties of this algorithm when the solution space $\scU\cap\scW$ is infinite due to insufficient sampling, or when it is empty when the acquired data is deviated by some error
\begin{equation}\label{data-proj-d}
w=P_\scV\' x+d.
\end{equation}
By combining knowledge from the POCS literature \cite{Bauschke94} and from functional analysis \cite{Kato95,Engl96}, we show that $x\up{n}$ tends to the least-squares solution of the equation $P_\scV u=w$ that is closest to the initial guess $x\up{0}$, whenever the equation $P_\scV u=w$ is theoretically well-posed in a sense that will be explained in the paper. Moreover, the algorithm possesses an inherent mechanism of regularization by semi-convergence \cite{Engl96}, which controls  the amplification of reconstruction errors due to data noise. What allows the extraction of all these properties is the derivation of various {\em equivalent} forms to the POCS iteration of \eqref{POCS-abs}, which we show in Table \ref{tab}. When the data $w$ is discrete, which is by default the case of sampling, the table shows additionally in (v) a systematic discrete-time implementation of the iteration, followed by a single continuous-time conversion at the final iteration. This does not require the use of a pre-existing orthogonal basis of $\scU$ (such as the sinc basis in the traditional framework of Shannon's sampling theory) and thus allows sophisticated input spaces that need not be separable\footnote
{When the input space $\scU$ is not separable, note that discrete sampling is by necessity insufficient. Nonetheless, the mentioned property of optimal input reconstruction from the available data by the POCS iteration remains valid.}.
\begin{table*}\label{tab}
\small
\renewcommand{\arraystretch}{2}
\begin{center}
\caption{Equivalent forms of POCS iteration}
\begin{tabular}{|c|c|c|c|c|}
\cline{2-5}
\multicolumn{1}{c|}{} & version & iteration, $n\geq0$  & eq.  & definitions and assumptions\\
\hline\hline
(i)  & original & $x\up{n+1}=P_\scU P_\scW^{}\hspace{0.5pt}x\up{n}$ & \eqref{POCS-abs} & \raisebox{-3mm}{$\scW:=\big\{u\in\scH:P_\scV u=w\big\}$}\\[-1ex]
\cline{1-4}
(ii) &\! $w$-data explicit\! & \!\!$x\up{n+1}=x\up{n}+P_\scU(w-P_\scV x\up{n})$\!\! & \!\eqref{POCS-proj}\! & \raisebox{-2mm}{$w\in\scV\quad,\quad
Q:\begin{array}[t]{rcl}\scU\!\!&\!\!\to\!\!&\!\!\scV\\[-2.5ex]
u \!\! & \!\! \mapsto \!\! & \!\! P_\scV\hspace{1pt} u\end{array}$}\\[-3.5ex]
\cline{1-4}
\!(iii)\!  & Landweber iteration & \!$x\up{n+1}=x\up{n}+Q^*(w-Q x\up{n})$\! & \!\eqref{Landweber}\! &
\\
\hline
\!(iv)\!  & discrete data & \!$x\up{n+1}=x\up{n}+V^*(\vw-V x\up{n})$\! & \!\eqref{Landweber2}\! &
$(v_k)\inZ$ = orthogonal basis of $\scV$\\
\cline{1-4}\\[-6.5ex]
\raisebox{-2mm}{\!(v)\!} & \!discrete-time iteration\! & \!\!$\begin{array}[t]{rcl}
\vc\up{n+1}&\!\!\!\!\!=\!\!\!\!&\vc\up{n}+ (\vw_0- VV^* \vc\up{n})\\[-2ex]
x\up{n}&\!\!\!\!\!=\!\!\!\!&x\up{0}+V^* \vc\up{n}\end{array}$\! & \raisebox{-2mm}{\!\eqref{sys}\!\!} & \raisebox{2mm}{$\begin{array}[t]{l}
\vw:=(\langle w,v_k\rangle)\inZ\\[-2ex]
\vw_0:=\vw-Vx\up{0}\end{array},~~
V:\begin{array}[t]{rcl}\scU\!\!&\!\!\to\!\!&\!\!\scD\\[-2.5ex]
u \!\! & \!\! \mapsto \!\! & \!\! (\langle u,v_k\rangle)\inZ\hspace{1pt}\end{array}$}
\\[-5ex]
& continuous-time conversion & & & \raisebox{-1mm}{$\vc\up{0}=0$} \\
\hline
\end{tabular}
\end{center}
\end{table*}

The article is organized as follows. To motivate our proposed framework, we first provide in Section \ref{sec:examples} examples of data acquisition that take the form of \eqref{data-proj-d}. The case of major importance in practice is when samples of the type
\begin{equation}\label{samp-gen-d}
\w_k:=\langle x,v_k\rangle+\d_k,\qquad k\in\Z
\end{equation}
are acquired from the input $x$, where $\Z$ is some finite or infinite index set,  $\langle\cdot,\cdot\rangle$ is the inner product of some ambient space $\scH$ containing $\scU$, $(v_k)\inZ$ is an {\em orthogonal} family of $\scH$, and $(\d_k)\inZ$ is some error sequence. Data knowledge of the form \eqref{samp-gen-d} can be equivalently expressed as \eqref{data-proj-d} by defining $\scV$ as the closed linear span of $(v_k)\inZ$ and $\vw$ as the vector of $\scV$ with components $(\w_k/\|v_k\|^2)\inZ$ in the basis $(v_k)\inZ$. This is the special case of generalized sampling introduced in \cite{eldar2005general}, where the sampling kernel functions are orthogonal in a space larger than the input space.  We then give explicit examples of this situation which include in particular the encoding schemes studied by Lazar and Tóth in \cite{Lazar04}.
In Section \ref{sec:POCS}, we return to the general assumption of \eqref{data-proj-d} and present the complete theory behind the convergence properties of iteration \eqref{POCS-abs} mentioned above. In Section \ref{sec:ortho-samp}, we derive the special forms that the POCS iteration yields when the original data is of the type of \eqref{samp-gen-d}, including the discrete-time version of the iteration (items (iv) and (v) of Table \ref{tab}). As a first application, we show in Section \ref{sec:MIMO} how the complex multi-channel time-encoding system introduced in \cite{Adam20,Adam20b,Adam21} can be reduced to the sampling framework of \eqref{samp-gen-d}. As an outcome, this simplifies the analysis of this system, while uncovering unexplored convergence properties of the reconstruction algorithms proposed in these references, especially in the conditions of noisy or insufficient sampling, up to some algorithmic rectifications in certain cases. This application is also an opportunity to generalize the sampling principle by Lebesgue integration over $\RR$ that governs the encoding system of Lazar and Tóth, to integration of signals over generalized measurable spaces. As a second application, we provide in Section \ref{sec:sob}, we a new interpretation to an algorithm introduced in \cite{Grochenig92b} for the bandlimited interpolation of nonuniform point samples. With such samples, \eqref{samp-gen-d} with orthogonal kernels can never be achieved in $L^2(\RR)$ (except in Nyquist rate sampling). However, nonuniform point samples can be formalized as \eqref{samp-gen-d} when choosing $\scH$ to be a Sobolev space. Like in the first application, we uncover in this way new convergence properties to an existing algorithm in similar conditions of non-ideal sampling. In these two applications, we moreover apply our rigorous method of iteration discretization on algorithms that have been left in the form of continuous-time operations.

\section{Examples of orthogonally-projected data}\label{sec:examples}

In this section, we give examples of data acquisition schemes that consist in recording from an input $x\in\scU$ data of the form \eqref{data-proj} with inherent errors as in \eqref{data-proj-d}. This takes place in an ambient Hilbert space $\scH$ equipped with an inner product $\langle\cdot,\cdot\rangle$ and a norm $\|\cdot\|$, with respect to which $\scU$ and $\scV$ are closed. Whenever $\scH=L^2(\RR)$, we will use the explicit notation of $\langle\cdot,\cdot\rangle_2$ and $\|\cdot\|_2$, We will denote by $\scB$ the space of bandlimited functions in $L^2(\RR)$ of Nyquist period 1. Any bandlimited signal can be thought of as an element of $\scB$ up to a change of time unit.

\subsection{Bandlimited extrapolation}\label{subsec:band-extra}

We first connect \eqref{data-proj} to the most classic problem of bandlimited extrapolation. Suppose that a bandlimited signal $x\in\scB$ is only known by its values $x(t)$ in a certain time interval $I$. Extrapolating $x(t)$ from $I$ can be expressed as recovering $x$ from $w$ in \eqref{data-proj} where
$$\scH:=L^2(\RR),\quad\scU:=\scB,\quad\scV:=L^2(I),\quad w:=1_I.\'x,$$
$1_I$ denotes the indicator function of the interval $I$ and $L^2(I)$ is seen as the subspace of $L^2(\RR)$ of functions supported by $I$. It is easy to verify that $1_I\'u=P_\scV\' u$ for all $u\in\scH$. This is the context of the Papoulis-Gerchberg algorithm mentioned in the introduction.

\subsection{Nonuniform sampling with orthogonal kernels}\label{subsec:ortho-samp}

Assume that samples of the type \eqref{samp-gen-d} are acquired from an input $x\in\scU\subset\scH$, where $(v_k)\inZ$ is an orthogonal family of $\scH$ and $(\d_k)\inZ$ is a sequences of errors. Then, it is easy to see that the generic data form of $w=P_\scV x+d$ we mentioned in \eqref{data-proj-d} is achieved with
\begin{subequations}\label{Vw}
\begin{eqnarray}
&\scV:=\overline\spn(v_k)\inZ,\label{scV}\\[0.5ex]
&w:=\smallsum{k\in\Z}\w_k\'v_k/\|v_k\|^2\quad\mbox{and}\quad
d:=\smallsum{k\in\Z}\d_k\'v_k/\|v_k\|^2.&\quad\label{wd}
\end{eqnarray}
\end{subequations}
Indeed, as $(v_k/\|v_k\|)\inZ$ is an orthonormal basis of $\scV$, we have
\begin{equation}\label{PV-samp-gen}
\forall u\in\scH,\qquad P_\scV\'u=\smallsum{k\in\Z}\langle u,v_k\rangle\'v_k/\|v_k\|^2.
\end{equation}
Then, by injecting \eqref{samp-gen-d} into the above expression of $w$, we obtain $w=P_\scV x+d$ as in \eqref{data-proj-d}. For the proper definition of $d$, we will assume that $(\d_k/\|v_k\|)\inZ$ is square-summable. As a result, $d\in\scV$.

\subsection{Nonuniform sampling by integration}\label{subsec:int}

We give here a concrete application of Section \ref{subsec:ortho-samp}.
Suppose that one acquires from an input $x\in\scB$ a sequence of integral values
\begin{equation}\label{samp-int}
\w_k:=\int_{t_{k-1}}^{t_k}x(t)\'f_k(t)\'\dif t+\d_k,\qquad k\in\Z
\end{equation}
where $\Z$ is a set of consecutive integers,  $(t_k)_{k\in\Z}$ is some increasing sequence of instants and $(f_k)\inZ$ is a known sequence of functions with sufficient regularity. Then \eqref{samp-int} is the  particular case of \eqref{samp-gen-d} where
\begin{eqnarray*}
&\scH:=L^2(\RR),\qquad\scU:=\scB,\qquad v_k:=1_{I_k}\'f_k,~~ k\in\Z\\
\mbox{with}&I_k:=[t_{k-1},t_k).\qquad k\in\Z.&\qquad
\end{eqnarray*}
The family $(v_k)\inZ$ is indeed orthogonal in $\scH$ since its functions have disjoint supports.

This type of sampling has attracted attention in event-based sampling since the design of a simple and robust analog circuit in \cite{Lazar04} to acquire samples of the type \eqref{samp-int} with $f_k=1$. As $v_k=1_{I_k}$ in this case, it follows from \eqref{scV} that $\scV$ is nothing but the space of piecewise constant functions in the partition of intervals $(I_k)\inZ$. A sampling model that is closer to the realistic behavior of practical integrating circuits is to include the effect of leakage by using the function $f_k(t):=e^{-\alpha(t-t_{k-1})}$ for some constant $\alpha>0$ \cite{stata1967operational,norsworthy1996delta,nahvi2017schaum}. This type of sampling is also involved in the most popular model of neurons, called  integrate-and-fire neurons \cite{stein1965theoretical}. In general, one can include any design of function $f_k(t)$ to optimize the matching between the sampling model and the actual system (electronic or biological). For example, the refractory effect of an integrate-and-fire neuron considered in \cite{Lazar04b} amounts here to choosing $f_k:=1_{I'_k}$ where $I'_k:=[t_{k-1}\!+\delta,t_k)\varsubsetneq I_k$.

\subsection{Nonuniform point sampling}\label{subsec:Groch}

Even traditional point sampling can be formalized as acquiring data of the form \eqref{data-proj}, up to working with some non-standard Hilbert space.
Suppose that one acquires from an input $x\in\scB$ point samples $(x(t_k))\inZ$ where $\Z$ is a set of consecutive integers and  $(t_k)_{k\in\Z}$ is some increasing sequence of instants. We have the knowledge to form the function $w(t)$ that linearly interpolates the points $(t_k,x(t_k))\inZ$. Let $\scV$ be the space of functions that are linear in every interval $[t_{k-1},t_k]$ with square-summable derivatives over $\RR$. It can be shown that $w$ is equivalently the element of $\scV$ that minimizes $\|w'-x'\|_2$. We will see in Section \ref{sec:sob} that this can be formalized as $w=P_\scV x$ in the homogeneous Sobolev space \cite{grafakos2014modern,devore1993constructive}
\begin{align}
&\scH=\sob:=\label{sob}\\
&\Big\{u(t):u \mbox{ is absolutely continuous on $\RR$ and } u'\in L^2(\RR)\Big\}.\nonumber
\end{align}

\section{Reconstruction by POCS}\label{sec:POCS}

In this section, we review in theoretical depth the properties of iteration \eqref{POCS-abs}, which we will call in short the POCS iteration. This will be a combination of results from the literature and our own synthesis.

\subsection{Properties of convergence from POCS literature}

Iterations of the type  \eqref{POCS-abs} were first studied in \cite{Neumann1949} in the general case where $\scU$ and $\scW$ are closed convex sets. For any such set $\scS$, $P_\scS u$ is the element of $\scS$ that is closest to $u$ in norm. With the additional assumption that $\scU$ and $\scW$ are affine subspaces, which is the case of this paper, it was shown that $x\up{n}$ converges in norm to
\begin{equation}\label{lim0}
x\up{\infty}=P_{\scU\cap{\scW}}\'x\up{0}.
\end{equation}
In other words, $x\up{\infty}$ is the point of $\scU\cap\scW$ that is closest to the initial guess $x\up{0}$ in norm.
The difficult situation of interest in this paper is when $\scU\cap{\scW}=\emptyset$.  Theorem 4.1 of \cite{Bauschke94} states that $(x\up{n})\n0$ remains convergent as long as the distance of $u$ to $\scW$, defined by
\begin{equation}\label{duW}
d(u,{\scW}):=\inf_{v\in\scW}\|u{-}v\|=\|u-P_\scW u\|,
\end{equation}
can be minimized within $\scU$. In this case, the limit of $x\up{\infty}$ is a point of
\begin{equation}\label{UW}
\scU_\scW:=\Big\{u^*\in\scU:d( u^*,\scW)=\inf_{u\in\scU}d( u,\scW)\Big\}.
\end{equation}
The complete behavior of $x\up{n}$ is given by the following theorem.
\line
\begin{theorem}[\cite{Bauschke94}]\label{theo:Bauschke}
Let $(x\up{n})\n0$ be recursively  obtained by the POCS iteration of \eqref{POCS-abs} for some initial iterate $x\up{0}\in\scH$.
\begin{description}
\item If $\scU_\scW\neq\emptyset$, $x\up{n}$ converges of limit
\begin{equation}\label{lim1}
x\up{\infty}=P_{\scU_\scW}x\up{0}.
\end{equation}
\item If $\scU_\scW=\emptyset$, $\|x\up{n}\|$ tends to $\infty$.
\end{description}
\end{theorem}
\ppnoi
When $\scU\cap{\scW}\neq\emptyset$, it can be seen that $\scU_\scW=\scU\cap{\scW}$ from the basic fact that $d( u,{\scW})=0$ if and only if $u\in{\scW}$.

\subsection{Dependence of $\scU_\scW$ with data $w$}

As mentioned in the introduction, the space $\scW$ of interest is specifically defined by
\begin{equation}\label{W-proj}
\scW:=\big\{u\in\scH:P_\scV\' u=w\big\}.
\end{equation}
When $w$ is given by \eqref{data-proj}, $x\in\scU\cap\scW=\scU_\scW$ as seen above. When $w$ uniquely characterizes a signal in $\scU$, then $\scU\cap\scW$ is singleton. Then, \eqref{lim0} or equivalently \eqref{lim1} leads to the perfect reconstruction of $x$. But when $w$ is corrupted by noise as in \eqref{data-proj-d}, we end up with two questions:
\begin{enumerate}
\item When $\scU_\scW\neq\emptyset$, what is the meaning of the convergence limit of  \eqref{lim1}?
\item What is the condition for iteration \eqref{POCS-abs} to be stable with respect to data errors, i.e., for $\scU_\scW$ to remain non-empty regardless of the data deviation $d$, according to Theorem \ref{theo:Bauschke}?
\end{enumerate}
Motivated by the result of Section \ref{subsec:ortho-samp}, we will systematically assume that $d\in\scV$, which implies that $w\in\scV$.

For any of these questions, what needs to be determined is the dependence of $\scU_\scW$ with $w$.
Given that $w\in\scV$, we have the following equivalences:  $u\in\scW$ $\Leftrightarrow$ $P_\scV u=w=P_\scV w$ $\Leftrightarrow$ $P_\scV(u{-}w)=0$ $\Leftrightarrow$ $u{-}w\in\scV^\perp$ $\Leftrightarrow$ $u\in\scV^\perp\!+w$. Thus,
\begin{equation}\label{W-decomp}
\scW=\scV^\perp\!+w.
\end{equation}
This gives us access to a more explicit expression of $P_\scW$.
\line
\begin{proposition}\label{lem:aff-proj}
\begin{align}\label{PW}
\forall u\in\scH,\qquad P_\scW u= u+( w-P_\scV\' u).
\end{align}
\end{proposition}

\begin{IEEEproof}
By space translation, $v=P_\scW u$ $\Leftrightarrow$ $v{-}w=P_{\scV^\perp}(u{-}w)$ $\Leftrightarrow$ $v=P_{\scV^\perp}(u{-}w)+w=u-P_\scV u+w$ since $w\in\scV$.
\end{IEEEproof}
\ppnoi
We thus find that $d(u,\scW)=\|P_\scW u-u\|=\|w-P_\scV u\|$. We can then write
\begin{equation}\label{UW2}
\scU_\scW=\Big\{u^*\in\scU:\| w-P_\scV u^*\|^2=
\displaystyle\inf_{u\in\scU}\| w-P_\scV u\|^2\Big\}.
\end{equation}
Thus, $\scU_\scW$ is the set of least-squares solutions of the equation $P_\scV u=w$. This leads to the following qualitative interpretation of Theorem \ref{theo:Bauschke}.
\line
\begin{proposition}
When $\scU_\scW\neq0$, the limit \eqref{lim1} of iteration \eqref{POCS-abs} is the least-squares solution of the equation $P_\scV u=w$ that is closest in norm to the initial guess $x\up{0}$.
\end{proposition}
\ppnoi
This answers our first question. The second question is addressed in the next two subsections.

\subsection{General result on least-squares solutions}\label{subsec:gen-Ls}

The second question amounts to finding the condition for $\scU_\scW$ in \eqref{UW2} to be non-empty for all $w\in\scV$.
It will be convenient to study this question in a more abstract setting. For any bounded linear operator $Q$ between Hilbert spaces
\begin{equation}\label{Qgen}
Q:(\scU,\langle\cdot,\cdot\rangle_\scU)~\rightarrow~(\scV,\langle\cdot,\cdot\rangle_\scV)
\end{equation}
the problem is to find the condition for the set
\begin{equation}\label{SQw}
\scU_{Q,w}:=\Big\{u^*\in\scU:\| w-Q u^*\|_\scV^2=
\displaystyle\inf_{u\in\scU}\| w-Q u\|_\scV^2\Big\}
\end{equation}
to be non-empty for any $w\in\scV$. The answer relies on the following lemma.
\line
\begin{lemma}\label{lem1}
Let $\scR$ be a linear subspace of ${\scV}$ and $w$ be a fixed element of ${\scV}$. Then, $\|w-v\|$ has a minimizer $v^*$ in $\scR$ if and only if $w\in\scR\oplus\scR^\perp$.
\end{lemma}
\line
\begin{IEEEproof}
Let $\overline\scR$ be the closure of $\scR$ and $w=w\subsmall\|+w\subsmall\perp$ be the decomposition of $w$ in ${\overline\scR}{\times}{\overline\scR}^\perp$. For any $v\in\scR$, $\|w{-}v\|^2=\|w\subsmall\|{-}v\|^2+\|w\subsmall\perp\|^2$. The infimum of $\|w\subsmall\|{-}v\|$ for $v\in\scR$ is 0, but can be attained in $\scR$ when and only when $w\subsmall\|\in\scR$. Thus, $\|w-v\|$ has a minimizer in $\scR$ if and only if $w\in \scR\oplus{\overline\scR}^\perp$. This completes the proof since ${\overline\scR}^\perp=\scR^\perp$.
\end{IEEEproof}
\ppnoi
We then have the following equivalences.
\line
\begin{theorem}\label{theo:closed-range}
The following statements are equivalent:
\begin{enumerate}[label=(\roman*)]
\item $\scU_{Q,w}\neq\emptyset$ for all $w\in\scV$,\vspace{0.5mm}
\item $\calR(Q)\oplus\calR(Q)^\perp=\scV$,\vspace{0.5mm}
\item $\calR(Q)$ is closed in $\scV$,\vspace{0.5mm}
\item $\gamma(Q)>0$, where
\begin{equation}\label{gamma}
\gamma(Q):=\inf\Big\{\|Qu\|_\scV:u\in\calN(Q)^\perp,\|u\|_\scU=1\Big\}
\end{equation}
and $\calN(Q)$ is the null space of $Q$.
 \end{enumerate}
\end{theorem}
\line
\begin{IEEEproof}
(i) $\Leftrightarrow$ (ii): From \eqref{SQw}, $\scU_{Q,w}$ is non-empty if and only if $\|w-v\|^2$ has a minimizer $v^*$ in the range $\calR(Q)$ of $Q$. By Lemma \ref{lem1} with $\scR:=\calR(Q)$, this is equivalent to $w\in\calR(Q)\oplus\calR(Q)^\perp$. This proves that (i) $\Leftrightarrow$ (ii).
\line
(ii) $\Leftrightarrow$ (iii): Since $\overline{\calR(Q)}=(\calR(Q)^\perp)^\perp$ in $\scV$, then $\scV=\overline{\calR(Q)}\oplus\calR(Q)^\perp$. Therefore, (ii) $\Leftrightarrow$ ${\calR(Q)}=\overline{\calR(Q)}$ $\Leftrightarrow$ (iii).
\line
(iii) $\Leftrightarrow$ (iv): This is shown in Theorem 2.5 of \cite[chap.IV]{Kato95}.
\end{IEEEproof}
\ppnoi
The coefficient $\gamma(Q)$ is called the reduced minimum modulus of $Q$  \cite[chap.IV]{Kato95}.

\subsection{Stability of POCS iteration \eqref{POCS-abs}}

Returning to our original space setting where $\scU$ and $\scV$ are closed subspaces of $(\scH,\langle\cdot,\cdot\rangle)$, the set $\scU_\scW$ from \eqref{UW2} is equal to $\scU_{Q,w}$ in \eqref{SQw} with the operator $Q$ defined by
\begin{equation}\label{Q}
\begin{array}[t]{rcl}
Q: \quad \scU& \rightarrow &\scV\\
 u& \mapsto &P_{\scV}u
\end{array}.
\end{equation}
As an answer to our second question, it follows from Theorem \ref{theo:closed-range} that iteration \eqref{POCS-abs} is stable if and only if $\calR(Q)=P_\scV(\scU)$ is closed. Note that this is automatically satisfied when either $\scU$ or $\scV$ is of finite dimension. This is because $P_\scV(\scU)$ is of finite dimension and hence closed. Therefore, the stability of  iteration \eqref{POCS-abs} is never an issue in practical sampling which consists in  acquiring a finite number of samples of the type \eqref{samp-gen-d} and hence implies  a finite-dimensional space $\scV$.

However, stability becomes an issue when dealing with the theoretical question of perfect reconstruction of signals in a space $\scU$ of infinite dimension (such as bandlimited signals in $L^2(\RR)$), since this is possible only when $\scV$ is also infinite-dimensional. In this case, we derive the condition of stability using the criterion that $\gamma(Q)>0$ from Theorem \ref{theo:closed-range}. Let us give a more explicit expression of $\gamma(Q)$ in \eqref{gamma} given our assumption of \eqref{Q}. A vector $u$ is in  $\calN(Q)$ if and only if $u\in\scU$ and $P_\scV u=0$, which is equivalent to $u\in\scI:=\scU\cap\scV^\perp.$
As $\calN(Q)^\perp\subset\scU$, then $\calN(Q)^\perp=\scU\cap\scI^\perp$. Thus, $\gamma(Q)$ yields the explicit description
\begin{equation}\label{gamma-explicit}
\gamma(Q)=\inf\Big\{\|P_{\scV} u\|:u\in\scU\cap\scI^\perp,\|u\|=1\Big\}.
\end{equation}
It can be verified that $\gamma(Q)$ is the principal angle between $\scU$ and $\scV^\perp$ (see \cite{Deutsch1995} for the definition of space angle). Now, the angle between subspaces in infinite dimension is a difficult topic that requires advanced research on a case-by-case basis. This question is however of pure theoretical interest. We address in the next subsections an issue of higher priority to practitioners on the numerical side.

\subsection{Landweber iteration interpretation}\label{subsec:gradient}

More properties on the numerical behavior of iteration \eqref{POCS-abs} can be obtained by deriving equivalent forms of it. It is clear from \eqref{POCS-abs} that $x\up{n}\in\scU$ for all $n\geq1$. Let us assume that $x\up{0}$ is chosen in $\scU$ as well, so that $P_\scU x\up{n}=x\up{n}$ for all $n\geq0$. By injecting \eqref{PW} into \eqref{POCS-abs}, we obtain
\begin{equation}\label{POCS-proj}
x\up{n+1}= x\up{n}+\'P_\scU(w-P_\scV x\up{n}),\qquad n\geq0.
\end{equation}
One can easily see that the adjoint of $Q$ is the operator
\begin{equation}\label{Q*}
\begin{array}[t]{rcl}
Q^*: \quad \scV & \rightarrow &\scU\\
v& \mapsto &P_\scU v
\end{array}.
\end{equation}
Indeed, since orthogonal projections are self-adjoint, we have for all $u\in \scU$ and $ v\in \scV$ that $\langle Qu,v\rangle=\langle P_{\scV}u,v\rangle=\langle u,P_{\scV}v\rangle=\langle u,v\rangle=\langle P_\scU u,v\rangle=\langle u,P_\scU v\rangle$.
Then, iteration \eqref{POCS-proj} yields the form of
\begin{equation}\label{Landweber}
x\up{n+1}= x\up{n}+\'Q^*(w-Q x\up{n}),\qquad n\geq0.
\end{equation}
This is a Landweber iteration \cite{Landweber51}. An outstanding consequence is addressed in the next subsection.

\subsection{Regularization by semi-convergence}\label{subsec:semiconv-regul}

Given our definition of stability, we saw that iteration \eqref{POCS-abs} is theoretically convergent regardless of data errors when $\gamma(Q)>0$ with the operator $Q$ of \eqref{Q}. In practice however, this condition is insufficient for the iteration to be numerically well-behaved. One encounters problems as soon as $\gamma(Q)/||Q\|$ is of the order of the machine precision or less, where $\|Q\|$ is the operator norm of $Q$. The term of ``ill-posed problem'' is often used in practice to imply this situation \cite{Strohmer00}. Remarkably, the Landweber iteration possesses an inherent mechanism to mitigate this issue through semi-convergence: it naturally regularizes the reconstruction  estimate when the iteration is stopped early enough. While extensive theories have been developed on this type of regularization \cite{Engl96}, we will give here only a few elements of it for qualitative understanding.

Assume that $\|Q\|\leq1$ (as is the case in \eqref{Q} since orthogonal projections are non-expansive) and  $Q$ is compact, which is automatically realized when $\scV$ is of finite dimension. Then $Q$ yields a singular system $(\varphi_i,\psi_i,\sigma_i)_{i\geq1}$, which satisfies the following properties: $(\varphi_i)_{i\geq1}$ and $(\psi_i)_{i\geq1}$ are orthonormal families in $\scU$ and $\scV$, respectively, $(\sigma_i)_{i\geq1}$ is a non-increasing sequence of non-negative numbers, $Q\varphi_i=\sigma_i\psi_i$ and $Q\psi_i=\sigma_i\varphi_i$ for all $i\geq1$. Using the notation $u_i:=\langle u,\varphi_i\rangle_\scU$ and $v_i:=\langle v,\psi_i\rangle_\scV$ for any $u\in\scU$, $v\in\scV$ and each $i\geq1$, one finds from \eqref{Landweber} that
$$x_i\up{n+1}=x_i\up{n}+\sigma_i(w_i-\sigma_i x_i\up{n}),\qquad n\geq0.$$
Let us write $w=Qx+d$ where $d$ is the data error, and let $e\up{n}:=x\up{n}-x$ be the resulting error of the $n$th iterate. Since $w_i=\sigma_i x_i+d_i$, subtracting $x_i$ from the above equation yields
$$e_i\up{n+1}=\mu_i\'e_i\up{n}+\sigma_i\'d_i\qquad\mbox{where}\qquad\mu_i:=1\!-\!\sigma_i^2.$$
Limiting ourselves to the singular directions in $\calN(Q)^\perp$, we have $\sigma_i>0$, and hence, $\mu_i\in[0,1)$ since $\sigma_i\in(0,\|Q\|]\subset(0,1]$. It can then be found by induction that
$$e_i\up{n}=\mu_i^n\'e_i\up{0}+(1\!-\!\mu_i^n)\'\frac{d_i}{\sigma_i},\qquad n\geq0.$$
The first term, which is the noise-free iteration error, monotonically tends to 0. Meanwhile, the second term, which is the noise contribution, on the contrary increases in magnitude with iteration. Roughly speaking, regularisation by semi-convergence consists in stopping the iteration when the two terms are globally of similar order. The critical singular directions are those along which $\sigma_i$ is small, as can be seen in the second term. However, note that the amplified error limit $d_i/\sigma_i$ also takes more iterations to be reached, as $\mu_i^n$ remains close to 1 for larger values of $n$. A complete theory is presented in \cite{Engl96}, showing that stopping the Landweber iteration at the right time leads to a regularized reconstruction that is optimal in a theoretical sense called order-optimality. A study of practical stopping rules is proposed in \cite{Elfving2007}.

\section{Sampling with orthogonal kernels}\label{sec:ortho-samp}

In this section, we focus on the important case where the space $\scV$ has a known orthogonal basis $(v_k)\inZ$. We saw in Section \ref{subsec:ortho-samp} that this situation is indirectly realized when the acquired data consists of samples $(\w_k)\inZ$ of the form
$$\w_k:=\langle x,v_k\rangle+\d_k,\qquad k\in\Z$$
as a reminder of \eqref{samp-gen-d}, for any given orthogonal family $(v_k)\inZ$ of the ambient space $\scH$.  In this case, $\scV$ and $w$ are defined by \eqref{Vw}. The task of this section is to express iteration \eqref{POCS-abs} and its properties in terms of the sequence of samples
$$\vw=(\w_k)\inZ$$
instead of $w$. An outstanding outcome will be the possibility to rigorously implement the iteration by means of discrete-time operations, regardless of the input space $\scU$.

\subsection{POCS algorithm}\label{subsec:POCS-discr}

Let us express iteration \eqref{POCS-abs} in terms of $(\w_k)\inZ$.
As $(v_k)\inZ$ are independent vectors (since they are orthogonal), it can be seen from \eqref{Vw} and \eqref{PV-samp-gen} that $P_\scV u=w$ if and only if $\langle u,v_k\rangle=\w_k$ for all $k\in\Z$. The space $\scW$ then becomes explicitly
$$\scW=\Big\{u\in\scH:~\forall k\in\Z,\langle u,v_k\rangle=\w_k\Big\}.$$
The explicit implementation of iteration \eqref{POCS-abs} in terms of $(\w_k)\inZ$ can be obtained by injecting \eqref{wd} and \eqref{PV-samp-gen} into the equivalent iteration of \eqref{POCS-proj}. After distribution of the projection $P_\scU$, this yields
\begin{equation}\label{POCS2}
x\up{n+1}= x\up{n}+P_\scU\Big(\smallsum{ k\in\Z}\big(\w_k-\langle  x\up{n}\!, v_k\rangle\big)\' v_k/\|v_k\|^2\Big).
\end{equation}
Note that $P_\scU$ need not be computed on any element of $\scH$, but only on the vectors $(v_k)\inZ$.

\subsection{Formulation in terms of sampling operator}

Iteration \eqref{POCS2} was established as an early expression of \eqref{POCS-abs} in terms of $\vw$. It however cumulates the shortcomings of preventing signal processing insight and of being impractical in implementation, as will be seen later on.
For a  conceptual understanding of this iteration, it is better to perform a more formal transformation of  \eqref{POCS-abs}.
By orthogonality of $(v_k)\inZ$, it follows from  \eqref{wd} that $\w_k=\langle w,v_k\rangle=\|v_k\|\'\langle w,\hat v_k\rangle$ where
$$\hat v_k:=v_k/\|v_k\|,\qquad k\in\Z.$$
We can then write that
\begin{eqnarray}
&\vw=DCw\label{ww}\\[1ex]
\mbox{where}
&C: \begin{array}[t]{rcl}
\scV& \rightarrow & \ell^2(\Z)\\
v& \mapsto &\big(\langle v,\hat v_k\rangle\big)\inZ
\end{array},\label{C}\\[0.5ex]
&D: \begin{array}[t]{rcl}
 \ell^2(\Z)& \rightarrow &\scD\\
(\c_k)\inZ& \mapsto &(\|v_k\|\c_k)\inZ
\end{array},\label{D}\\[0.5ex]
\mbox{and}&\scD:=\Big\{(\c_k)\inZ:\smallsum{k\in\Z}\c_k^2/\|v_k\|^2<\infty\Big\}.&\qquad\nonumber
\end{eqnarray}
The space $\scD$ is a Hilbert space equipped with the weighted norm $\|\cdot\|_\scD$ such that
\begin{equation}\label{Dnorm}
\|\vc\|_{\scD}^2=\smallsum{k\in\ZZ}\c_k^2/\|v_k\|^2,\qquad\vc\in\scD.
\end{equation}
The operator $C$ is unitary since $(\hat v_k)\inZ$ is an orthonormal basis of $\scV$. The operator $D$ is also unitary, this time because it is invertible and norm preserving, since $\|D\vc\|_\scD=\|\vc\|$ for all $\vc\in \ell^2(\Z)$. As $DC$ is unitary, $C^*D^*DC$ is the identity on $\scV$.
Iteration \eqref{POCS-proj} can then be rewritten as
$$x\up{n+1}= x\up{n}+Q^*C^*D^*DC(w-Q x\up{n}),\qquad n\geq0.$$
This leads to
\begin{equation}\label{Landweber2}
x\up{n}= x\up{n}+V^*(\vw-V x\up{n}),\qquad n\geq0
\end{equation}
where $V:=DCQ$. The next proposition gives the explicit descriptions of $V$ and $V^*$.
\line
\begin{proposition}
\begin{equation}
V:\begin{array}[t]{rcl}
\scU&\!\! \rightarrow \!\!&\scD\\
u&\!\! \mapsto\!\! &\big(\langle u,v_k\rangle\midlow{\scV}\big)\inZ
\end{array}\hspace{-2mm},\quad
V^*:\begin{array}[t]{rcl}
\scD&\!\! \rightarrow\!\! &\scU\\
\vc&\!\! \mapsto\!\! &\sum\limits_{k\in\Z}\c_k\' u_k
\end{array}\label{V}
 \end{equation}
 where $u_k:=P_\scU v_k/\|v_k\|^2$ for all $k\in\Z$.
\end{proposition}
\line
\begin{IEEEproof}
For all $u\in\scU$, $ V u=DCP_\scV u=(\langle P_\scV u,v_k\rangle)\inZ=(\langle u,v_k\rangle)\inZ$.  Let $\vc=(\c_k)\inZ\in\scD$. We have $V^*\vc=Q^*D^*C^*\vc=P_\scU D^{-1}C^{-1}\vc$.
Since $C^{-1}\vc=\sum\inZ\c_k \hat v_k$ and $\hat v_k=v_k/\|v_k\|$, then $D^{-1}C^{-1}\vc=\sum\inZ\c_k  v_k/\|v_k\|^2$. Finally, $V^*\vc=\sum\inZ\c_k u_k$ where $u_k=P_\scU v_k/\|v_k\|$.
\end{IEEEproof}
\ppnoi
We now have the direct relation $\vw=Vx+(\d_k)\inZ$. We call $V$ the {\em sampling operator}. One can verify that iteration \eqref{POCS2} is retrieved by injecting \eqref{V} into \eqref{Landweber2} and factoring out $P_\scU$. But the interesting contribution of \eqref{Landweber2} is its simple form as a Landweber iteration in terms of $\vw$.

\subsection{Convergence interpretation}

Theorem \ref{theo:Bauschke} still gives the convergence behavior of iteration \eqref{POCS-abs}, and hence of iteration \eqref{Landweber2}. One however wishes to have a description of the set $\scU_\scW$ of iteration limits in terms of $\vw$. Since $DC$ is unitary, $\|DCv\|_\scD=\|v\|$ for all $v\in\scV$. Then,
\begin{eqnarray}
&\forall u\in\scU,\qquad\|w-P_\scV u\|=\|w-Qu\|=\|\vw- V u\|_\scD\nonumber
\end{eqnarray}
so that $\scU_\scW$ in \eqref{UW2} becomes
$$\scU_\scW=\Big\{ u^*\in\scU:\| \vw-V u^*\|_\scD^2=
\displaystyle\inf_{u\in\scU}\| \vw-V u\|_\scD^2\Big\}.$$
Thus, $\scU_\scW$ is the set of least-squares solutions of the equation $Vu=\vw$, where the least-squares evaluation is  in the sense of the weighted norm $\|\cdot\|_\scD$.
By simple application of Theorem \ref{theo:Bauschke}, we can then state the following result.
\line
\begin{proposition}
Let $(x\up{n})\n0$ be recursively defined by \eqref{Landweber2} for some initial iterate $x\up{0}\in\scU$ and some sequence $\vw=(\w_k)\inZ\in\scD$.
\begin{description}
\item If $\scU_\scW\neq\emptyset$, $x\up{n}$ converges of limit
$x\up{\infty}=P_{\scU_\scW}x\up{0}$.
\item If $\scU_\scW=\emptyset$, $\|x\up{n}\|$ tends to $\infty$.
\end{description}
\end{proposition}
\ppnoi
When convergent, iteration \eqref{Landweber2}  tends to the least-squares solution of the equation $V u=\vw$ in $\scU$ which is closest to the initial iterate $x\up{0}$.

\subsection{Discrete-time implementation of iteration}\label{subsec:discrete}

It appears from \eqref{Landweber2} that $x\up{n+1}\!-x\up{n}$ remains in the range of $V^*$. So does $x\up{n}\!-x\up{0}$ for all $n\geq0$. Therefore, there must exist for each $n\geq0$ a discrete-time vector $\vc\up{n}\in\scD$ such that
\begin{equation}\label{uc}
x\up{n}=x\up{0}+V^*\vc\up{n},\qquad n\geq0.
\end{equation}
As $V^*$ may not be injective, the sequence $(\vc\up{n})_{\geq0}$ is not necessarily unique. Let us show that \eqref{uc} is at least achieved by the discrete-time sequence recursively defined by
\begin{align}\label{c-iter}
\vc\up{n+1}:= \vc\up{n}+  (\vw_0- VV^* \vc\up{n}),\qquad n\geq0
\end{align}
 where $\vw_0:=\vw-V x\up{0}$, starting from $\vc\up{0}=0$. As \eqref{uc} is trivially true at $n=0$, let us assume that it is true at some $n\geq0$. By applying $V^*$ to the members of \eqref{c-iter} and adding $x\up{0}$, we obtain
\begin{align*}
x\up{0}+V^* \vc\up{n+1}&=x\up{0}+V^* \vc\up{n}+V^*(\vw_0- VV^* \vc\up{n})\\
&= x\up{n}+V^*\big(\vw_0- V(x\up{n}\!-x\up{0})\big)\\
&= x\up{n}+V^*(\vw- V x\up{n})=x\up{n+1}.
\end{align*}
Thus, \eqref{uc} is true for all $n\geq0$. We have thus established the following result.
 \line
 \begin{proposition}
 For a given initial guess $x\up{0}\in\scU$, the sequence $(x\up{n})\n0$ recursively defined by Landweber iteration \eqref{Landweber2} is equivalently obtained by the system
\begin{subequations}\label{sys}
\begin{align}
\vc\up{n+1}&=\vc\up{n}+  (\vw_0- VV^* \vc\up{n})\label{sysa}\\
x\up{n}&=x\up{0}+V^* \vc\up{n}\label{sysb}
\end{align}
\end{subequations}
starting from $\vc\up{0}=0$, where $\vw_0:=\vw-V x\up{0}$.
 \end{proposition}
 \ppnoi
 If the targeted estimate of $x$ is the $m$th iterate $x\up{m}$, then one only needs to iterate $m$ times the discrete-time operation of \eqref{sysa}, then get $x\up{m}$ by executing \eqref{sysb} once at $n=m$. For storage and transmission, $x\up{m}$ may not even need to be computed. One may wait until the signal $x\up{m}$ needs to be output physically to perform \eqref{sysb} from $\vc\up{m}$ by means of analog circuits. In \eqref{sysa}, the whole complexity lies in the matrix-vector multiplication $(VV^* )\vc\up{n}$ .
\line
\begin{proposition}
$VV^*$ is the matrix of coefficients
\begin{equation}\label{VV*}
VV^*=\Big[\langle P_\scU v_\ell,v_k\rangle/\|v_\ell\|^2\Big]_{(k,\ell)\in\Z\times\Z}.
\end{equation}
\end{proposition}

\begin{IEEEproof}
For any given $\ell\in\Z$, let $\ve_\ell$ be the sequence in $\scD$ with 1 as the $\ell$th coordinate and 0 elsewhere. Then, $VV^*\ve_\ell=V u_\ell=\big(\langle u_\ell,v_k\rangle\big)_{k\in\Z}=\big(\langle P_\scU v_\ell,v_k\rangle/\|v_\ell\|^2\big)_{k\in\Z}$.
\end{IEEEproof}

\subsection{Discussion}

Signal processing algorithms often refer to continuous-time signals but are typically thought of as discrete-time algorithms via the Shannon sampling theorem under bandlimitation. This is for example the context of \cite{Lazar04,Adam20b,Adam21}. This could be applied to iteration \eqref{POCS2}.
This would however create a double summation in \eqref{POCS2}, as $x\up{n}$ in the inner product $\langle x\up{n},v_k\rangle$ would need to be expanded in terms of its components in this basis. Meanwhile, the discretization technique of \ref{subsec:discrete} solely results from the inherent discreteness of the acquired data $\vw$. In this context, the space $\scU$ need not even be separable (except when one aims at the perfect reconstruction of $x$ from $\vw$).

\section{Multi-channel time encoding}\label{sec:MIMO}

\begin{figure}
\centerline{\hbox{\scalebox{0.95}{\includegraphics{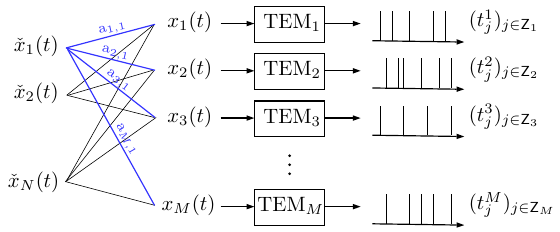}}}}
\caption{Multi-channel system of  time-encoding machines (TEM) from \cite{Adam20,Adam21}.}\label{fig:Karen}
\end{figure}
To illustrate the power of our framework, we show that a complex multi-channel time-encoding system recently introduced in \cite{Adam20,Adam20b,Adam21} and shown in Fig. \ref{fig:Karen} amounts to acquiring data of the form \eqref{samp-gen-d} from an input in a certain space $\scU$. According to Section \ref{sec:ortho-samp}, this allows us to perform the input reconstruction using the POCS iteration of \eqref{POCS-abs} with all the powerful convergence properties of Sections \ref{sec:POCS} and \ref{sec:ortho-samp}. For efficiency, we will directly provide the time-discretized version \eqref{sys} of the POCS iteration with the parameters that are specific to the presently considered system.

\subsection{Encoding system description and problem}\label{subsec:mult-descr}

Inspired by the modeling of spiking neural networks, the works of \cite{Adam20,Adam20b,Adam21} studied the encoding of a multichannel function
\begin{equation}\label{mult-x}
\bx=(x_1(t),{\cdots},x_M(t))
\end{equation}
by parallel time encoding of each channel component $x_i$. As an extension of single-channel time encoding \cite{Lazar04}, it is assumed that the encoder of the $i$th channel outputs from $x_i$  a sequence of integral values of the type
\begin{equation}\label{wij}
\w_{ij}:=\int_{t^i_{j-1}}^{t^i_j}x_i(t)\,\dif t+\d_{ij},\qquad j\in\Z_i
\end{equation}
where $\Z_i$ is some set of consecutive integers, $(t^i_j)_{j\in\Z_i}$ is some increasing sequence of instants, and $(\d_{ij})_{j\in\Z_i}$ is some sequence of acquisition errors. Meanwhile, it is assumed for each $t\in\RR$ that
\begin{equation}\label{xAx}
(x_1(t),{\cdots},x_M(t))=\A\,(\check x_1(t),{\cdots},\check x_N(t))
\end{equation}
where $\A$ is a full rank $M{\times}N$ matrix with $N\leq M$, the $M$-tuple and $N$-tuple are seen as column vectors, and $\check x_1,\cdots,\check x_N$ are source signals in $\scB$. By abuse of notation, we will write that
$$\bx=\A\cbx\qquad\mbox{where}\qquad \cbx=(\check x_1(t),{\cdots},\check x_M(t)).$$
The goal is to estimate $\bx$ from the sample sequence
\begin{eqnarray}
&\vw:=(\w_{ij})_{\ij\in\Z}\nonumber\\
\lefteqn{\mbox{where}}\nonumber\\
&\Z:=\big\{\ij:i\in\M\mbox{ and } j\in\Z_i\big\}\quad\mbox{and}\quad\M:=\{1,{\cdots},M\}.\nonumber
\end{eqnarray}
If $\bx$ can be reconstructed perfectly, then $\cbx$ is retrieved by the transformation
\begin{equation}\label{source-rec}
\cbx=\A\!^+\bx
\end{equation}
where $\A\!^+$ is the pseudo-inverse of $\A$.

\subsection{Space setting and data formalism}\label{subsec:set-approach}

The central point of our approach to this encoding system is to look at a multichannel signal
$\bu=\big(u_1(t),\cdots,u_M(t)\big)$ as a 2D function
\begin{equation}\label{2Dsignal}
\bu:\begin{array}[t]{ccl}
\M{\times}\RR&\!\to\!&\RR\\
(\ell,t)&\!\mapsto\!&\bu(\ell,t):=u_\ell(t)
\end{array}.
\end{equation}
To measure the error between multichannel signals, we will work in the Hilbert space $$\scH:=L^2(\M{\times}\RR,\mu)$$
where $\mu$ is the product of the counting measure on $\M$ and the Lebesgue measure on $\RR$.  This space is equipped with the inner product
\begin{align}
\langle \bu,\bv\rangle&:=\int_{\M\times\RR}\bu(\ell,t)\'\bv(\ell,t)\'\dif\mu(\ell,t)\nonumber\\
&=\smallsum{\ell\in\M}\int_\RR \bu(\ell,t)\'\bv(\ell,t)\'\dif t,\qquad \bu,\bv\in\scH.\label{mult-inprod}
\end{align}
This view allows a generalization of Section \ref{subsec:int} on the generation of orthogonal kernel functions. If $(\bv_k)\inZ$ is a family of functions of $\scH$ supported by disjoint subsets of $\M{\times}\RR$, then $(\bv_k)\inZ$ is automatically orthogonal in $\scH$. To formalize the encoding scheme of Section \ref{subsec:mult-descr}, consider the family $(\bv_{ij})_{\ij\in\Z}$ defined by
\begin{equation}\label{vij}
\bv_{ij}:=1_{\{i\}\times I_{ij}}\qquad\mbox{where}\qquad I_{ij}:=[t^i_{j-1},t^i_j).
\end{equation}
The sets $(\{i\}{\times}I_{ij})_{\ij\in\Z}$ are disjoint in $\M{\times}\RR$. So, $(\bv_{ij})_{\ij\in\Z}$ is orthogonal in $\scH$. Next, since $\bv_{ij}(i,t)=1_{I_{ij}}(t)$ while $\bv_{ij}(\ell,t)=0$ for all $\ell\neq i$,
$$\int_{t^i_{j-1}}^{t^i_j}\!\!\!x_i(t)\,\dif t=\int_{I_{ij}}\!\!\bx(i,t)\'\dif t=\int_\RR \bx(i,t)\'\bv_{ij}(i,t)\'\dif t=\langle \bx,\bv_{ij}\rangle$$
from \eqref{mult-inprod}. It then follows from \eqref{wij} that
$$\w_{ij}=\langle \bx,\bv_{ij}\rangle+\d_{ij},\qquad\ij\in\Z$$
which is of the form of \eqref{samp-gen-d}.

\subsection{Input space $\scU$ and projection}

Looking at $\bx$ as a 2D function of $\scH$ according to \eqref{2Dsignal}, relation \eqref{xAx} implies that $\bx$ must remain confined into some subspace $\scU$ of $\scH$. Specifically, for each $\ell\in\M$, $\bx(\ell,\cdot)=x_\ell\in\scB$. Meanwhile, for each $t\in\RR$, $\bx(\cdot,t)$ is the vector $(x_1(t),{\cdots},x_M(t))$, which belongs to the range $\mA$ of matrix $\A$
$$\rmA:=\calR(\A)\subset\RR^M.$$
Thus, $\bx$ belongs to the space
\begin{align}
\scU:=\Big\{\bu\in\scH:~ &\bu(\cdot,t)\in\rmA\mbox{ for all } t\in\RR \nonumber\\[-1.5ex]
\mbox{and }~&\bu(\ell,\cdot)\in\scB\mbox{ for all }\ell\in\M\Big\}.\label{U-fib}
\end{align}
By principle, we can then use iteration \eqref{POCS-abs} to estimate $\bx$ from $\vw$. An initial difficulty is the derivation of $P_\scU$. However, we recall from Section \ref{subsec:POCS-discr} that only $P_\scU \bv_{ij}$ needs to be derived for $\ij\in\Z$. What will facilitate this derivation is the fact that $\bv_{ij}$ is a separable function. We say that a function $\bv\in\scH$ is separable when there exist $\v\in\RR^M$ and $v\in L^2(\RR)$ such that
$$\forall(\ell,t)\in\M{\times}\RR,\qquad\bv(\ell,t)=\v(\ell)\,v(t)$$
where $\v(\ell)$ designates the $\ell$th coordinate of $\v$. We will use the notation
$$\bv=\v\otimes v.$$
It can be seen from \eqref{vij} that
\begin{equation}\label{vij2}
\bv_{ij}=\e_i\otimes 1_{I_{ij}},\qquad\ij\in\Z
\end{equation}
where $\e_i$ is the standard basis vector of $\RR^M$ with 1 in the $i$th coordinate and 0 elsewhere.
We prove the following result in Appendix \ref{app:sep}.
\line
\begin{theorem}\label{theo:sep}
Let $\scU$ be defined by \eqref{U-fib}. Then,
\begin{equation}\label{eq:sep}
\forall\v\in\RR^M\!,v\in L^2(\RR),\quad P_\scU(\v\otimes v)=P_\rmA \v\otimes P_\scB v.
\end{equation}
\end{theorem}
\ppnoi
By applying this to $\bv_{ij}$ in \eqref{vij2}, we obtain
\begin{equation}\label{PUvij}
P_\scU \bv_{ij}=P_\rmA\e_i\otimes P_\scB 1_{I_{ij}},\qquad\ij\in\Z.
\end{equation}
As an $M{\times}M$ matrix, note that
\begin{equation}\label{PA}
P_\rmA=\A\A\!^+.
\end{equation}

\subsection{POCS iteration and discrete-time implementation}\label{subsec:mult-discr}

We have thus gathered all the conditions of Section \ref{sec:ortho-samp} to construct the POCS iteration of \eqref{POCS-abs} for the estimation of $\bx$ from $\vw$, with all the properties of convergence established in Sections \ref{sec:POCS} and \ref{sec:ortho-samp}. Given the equivalence of the various versions of POCS iteration in Table \ref{tab}, we can directly focus on the implementation of the practical iteration-discretized version of \eqref{sys}.  Under the current notation, this iteration is
\vspace{-4mm}
\begin{subequations}\label{sys-mult}
\begin{align}
\vc\up{n+1}&=\vc\up{n}+  (\vw_0- VV^* \vc\up{n})\label{sys-multa}\\
 \bx\up{n}&= \bx\up{0}+V^* \vc\up{n}\label{sys-multb}
\end{align}
\end{subequations}
starting from $\vc\up{0}=0$, where
\begin{eqnarray}
&\hspace{-5mm}V:\begin{array}[t]{rcl}
\scU&\!\! \rightarrow \!\!&\scD\\
 \bu&\!\! \mapsto\!\! &\big(\langle  \bu, \bv_{ij}\rangle\big)_{\ij\in\Z}
\end{array}\hspace{-1mm},~
V^*:\begin{array}[t]{rcl}
\scD&\!\! \rightarrow\!\! &\scU\\
\vc&\!\! \mapsto\!\! &\sum\limits_{\ij\in\Z}\!\!\c_{ij}\'  \bu_{ij}
\end{array}\hspace{-1mm}\label{Vmult}\\
& \vw_0:=\vw-V \bx\up{0},\quad \bu_{ij}:=P_\scU \bv_{ij}/\| \bv_{ij}\|^2,\quad\ij\in\Z.\nonumber
\end{eqnarray}
The general matrix expression \eqref{VV*} of $VV^*$ takes the form of
\begin{equation}\nonumber
VV^*=\Big[\langle P_\scU \bv_{i'j'}, \bv_{ij}\rangle/\| \bv_{i'j'}\|^2\Big]_{\big(\ij,(i',j')\big)\in\Z\times\Z}.
\end{equation}
The entries of matrix $VV^*$ need to be pre-computed  before the iteration. The yield the following formulas.
\line
\begin{proposition}\label{prop:bv-inner}
For all $\ij,(i',j')\in\Z$,
\begin{align}\label{eq:bv-inner}
\big\langle P_\scU \bv_{i'j'}, \bv_{ij}\big\rangle&=(\e_i^\top\!P_\rmA\e_{i'})
\,\big\langle P_\scB 1_{I_{i'j'}},1_{I_{ij}}\big\rangle_2\\
\| \bv_{i'j'}\|^2&=|I_{i'j'}|\nonumber
\end{align}
where $|I|$ denotes the length of any given interval $I$.
\end{proposition}
\line
\begin{IEEEproof}
By applying \eqref{mult-inprod}, \eqref{vij}  and \eqref{PUvij}, we obtain
$$\big\langle P_\scU \bv_{i'j'}, \bv_{ij}\big\rangle=\ssum{\ell\in\M}(P_\rmA\e_{i'})(\ell)\'\e_i(\ell)\!\int_\RR\!(P_\scB 1_{I_{i'j'}}\!)(t)\'1_{I_{ij}}(t)\,\dif t.$$
This leads to \eqref{eq:bv-inner}. Meanwhile, it results from \eqref{vij2} and \eqref{mult-inprod} that $\| \bv_{i'j'}\|^2=\|1_{I_{i'j'}}\|_2^2=|I_{i'j'}|$.
\end{IEEEproof}
\ppnoi
The inner product $\big\langle P_\scB 1_{I_{i'j'}},1_{I_{ij}}\big\rangle_2$ involved in \eqref{eq:bv-inner} can be computed using the formula
\begin{align}
\big\langle P_\scB1_{[a,b]},1_{[c,d]}&\big\rangle_2=\label{Karen-discrete}\\
&\varphi(d-a)-\varphi(d-b)-\varphi(c-a)+\varphi(c-b)\nonumber\\
\mbox{where}\qquad\qquad&\textstyle\varphi(t):=\int\limits_0^t(t{-}\tau)\, \sinc(\tau)\,\dif\tau,\label{phi}
\end{align}
which we prove in Appendix \ref{app-discrete}. One simply needs to inject in \eqref{Karen-discrete} $(a,b,c,d)=(t^{i'}_{j'-1},t^{i'}_{j},t^i_{j-1},t^i_j)$.
Assuming a default high-resolution quantization of the time instants $t^i_j$, the needed values of $\varphi(t)$ can be precalculated and stored in a lookup table. Overall, the required computation for each entry of the matrix $VV^*$ is limited to 4 table lookups, 8 additions and 2 multiplications.

\subsection{Final continuous-time output}\label{subsec:output}

Once \eqref{sys-multa} has been iterated the desired number of times $m$, one can output the continuous-time multi-channel signal $ \bx\up{m}(t)$ from \eqref{sys-multb}. For that purpose, we need to know the explicit expression of $V^* \vc$ in terms of $\vc=(\c_{ij})_{\ij\in\Z}$. It follows from \eqref{Vmult} and \eqref{PUvij} that
\begin{align*}
V^* \vc&=\textstyle\sum_{\ij\in \Z}\c_{ij}\,P_\scU \bv_{ij}/\|\bv_{ij}\|^2\\
&=\textstyle\sum_{\ij\in \Z}\c_{ij}\,P_\rmA\e_i\otimes P_\scB 1_{I_{ij}}/|I_{ij}|.
\end{align*}
By grouping all terms indexed by $j$ for a fixed $i$, we obtain
\begin{align}
V^* \vc=&\,\textstyle\sum_{i\in \M}P_\rmA\e_i\otimes P_\scB\' R_i\'\vc\label{ci}\\[0.5ex]
\mbox{where}\qquad\qquad R_i\'\vc:=&\,\textstyle\sum_{j\in\Z_i}\c_{ij}\,1_{I_{ij}}/|I_{ij}|,\qquad i\in\M.\qquad\nonumber
\end{align}
The function $(R_i\'\vc)(t)$ is nothing but the piecewise constant function equal to $\c_{ij}/|I_{ij}|$ in $I_{ij}$ for each $j\in\Z_i$. This is produced by analog circuits using a zero-order hold. The signal $(P_\scB\' R_i\'\vc)(t)$ is then obtained by lowpass filtering.

\subsection{Estimation of source signal}

Once an estimate $\bx\up{m}$ of $\bx$ is obtained from \eqref{sys-mult}, it results from \eqref{source-rec} that $\cbx\up{m}:=\A\!^+\bx\up{m}$ is an estimate of $\cbx$. Let $\vc\up{m}$ be the $m$th iterate of \eqref{sys-multa} starting from $\vc\up{0}=0$. It then follows from \eqref{sys-multb} that
$$\cbx\up{m}=\A\!^+ \bx\up{0}+\A\!^+V^*\vc\up{m}.$$
Let $\a_i^+$ be the $i$th column vector of $\A^+$. It results from \eqref{ci} that
$$\A\!^+ V^*\vc\up{m}=\ssum{i\in \M}\A\!^+P_\rmA\e_i\otimes P_\scB\' R_i\'\vc\up{m}=\ssum{i\in \M}\a^+_i\otimes P_\scB\' R_i\'\vc\up{m}$$
since $\A\!^+ P_\rmA\e_i=\A\!^+\A\A\!^+\e_i=\A\!^+\e_i=\a^+_i$ as a general result of pseudo-inverse.

\subsection{Comparison with prior work}

The prior work of \cite{Adam20,Adam20b,Adam21} introduced the encoding system considered in this section for the main objective of perfect input reconstruction. The signal reconstruction methods proposed in  \cite{Adam20,Adam20b} were inspired by the POCS algorithm. The simple case where $\A$ is the $M{\times}1$ matrix $\big[1\cdots 1\big]^\top$ was studied in \cite{Adam20b}. Although the adopted reconstruction algorithm was a non-standard version of POCS iteration (which can only be rigorously justified as alternating a projection with an average of projections \cite{Combettes99}), it can be verified to be equivalent to the POCS iteration of \eqref{POCS2} after re-indexation of $(\w_k,\bv_k)$ as $(\w_{ij},\bv_{ij})$. As a result, the theory of Section \ref{sec:POCS} uncovers all the convergence properties of their algorithm, including in the cases of inconsistent or insufficient data. Moreover, we give in \eqref{sys-mult} the time-discretized form of the iteration for practical implementation, whose complexity all lies in a matrix-vector multiplication. The work of \cite{Adam20} considered the general case of a full-rank tall matrix $\A$. While perfect reconstruction was numerically demonstrated, the reconstruction method used was a deviation of the POCS algorithm. As a result, it lacked a comprehensive theoretical framework for convergence analysis and the potential to converge beyond the conditions required for perfect reconstruction. Meanwhile, our POCS algorithm and analytical results hold with such a general matrix $\A$.

The more complete and concise analysis of the multichannel time-encoding system has also been an opportunity to outline a generalized construction of orthogonal sampling kernel by integration. The principle of integrating a signal over disjoint parts of its domain can be generalized from one-dimensional signals (which is realized with ASDM or integrate-and-fire) to multi-dimensional functions defined on any measurable space.

\section{Bandlimited interpolation by iterative piecewise-linear corrections}\label{sec:sob}

We now wish to apply our approach to the traditional point sampling problem of bandlimited signals. Assume that samples of the type
$$\x_k:=x(t_k)+\e_k,\qquad k\in\Z$$
are acquired from an input $x\in\scB$, where $\Z$ is some index set of consecutive integers,  $(t_k)_{k\in\Z}$ is some increasing sequence of instants and $(\e_k)\inZ$ some sequence of errors. The classical way to view $\x_k$ as $\langle x,v_k\rangle$ in $L^2(\RR)$ is to take $v_k(t)=\sinc(t{-}t_k)$. However, such a family of functions $(v_k)\inZ$ is orthogonal only when $t_k\!-t_\ell\in\ZZ$ for all $k\neq\ell$. An alternative attempt is to take $v_k(t)=\delta(t{-}t_k)$, thinking that these functions have disjoint supports regardless of the increasing sequence $(t_k)\inZ$. However, the Dirac impulse is not square-summable and hence does not fit in $L^2(\RR)$.

It appears that an iteration previously designed by Gr\"ochenig in \cite{Grochenig92b} for the bandlimited interpolation of point samples yields the POCS iteration form of (ii) in Table \ref{tab}, provided that the ambient space $\scH$ be the Sobolev space $\sob$ defined in \eqref{sob}. After presenting this iteration in its original form, we will justify our observation, thus revealing all the convergence properties of Section \ref{sec:POCS} in $\sob$. We will finally give some numerical results.

\subsection{Gr\"ochenig's algorithm}

Gr\"ochenig's method consists in iterating in $\scB$
\begin{equation}\label{groch-alg0}
x\up{n+1}=x\up{n}+P_\scB  L(\vx-S x\up{n}),\qquad n\geq0
\end{equation}
where $\vx=(\x_k)\inZZ$,  $Su:=(u(t_k))\inZZ$ and  $L\vc$ is the linear interpolation of the points $(t_k,\c_k)\inZZ$ for any sequence of values $\vc=(\c_k)\inZZ$. It was shown in \cite{Grochenig92b} that $x\up{n}$ linearly tends to $x$ in $L^2$-norm under the conditions that
\begin{enumerate}
\item $\lim_{k\rightarrow\pm\infty}t_k=\pm\infty$,
\item $\Delta:=\sup\inZZ\Delta t_k<1$ where $\Delta t_k:=t_k\!-t_{k-1}$,
\item $\e_k=0$ for all $k\in\ZZ$.
\end{enumerate}
Under the first condition alone, we are going to show that iteration \eqref{groch-alg0} is a POCS iteration in $\sob$.

\subsection{POCS iteration in Sobolev space $\sob$}\label{subsec:Linsob}

To imitate the presentation of Table \ref{tab}\'(ii), we can rewrite \eqref{groch-alg0} as
\begin{subequations}\label{groch-proj}
\begin{eqnarray}
\hspace{-4mm}&x\up{n+1}= x\up{n}+\'P_\scB(w-LS x\up{n}),& n\geq0\label{groch-proja}\\[0.5ex]
\hspace{-14mm}{\mbox{where}}&\hspace{10mm}w:=L\vx.\label{groch-projb}
\end{eqnarray}
\end{subequations}
We are going to see that $LS$ is an orthogonal projection in
$$\scH:=\sob.$$
In the definition of this space in \eqref{sob}, the absolute continuity on $\RR$  is in the local sense.  This means  that $u$ is differentiable almost everywhere with a locally integrable derivative $u'$ that satisfies the identity
$u(b)=u(a)+\int_a^b u'(t)\,\dif t$ for any $a\leq b$ \cite[\S11.4.6]{berberian2012first}.
This space is equipped with the inner product and the norm defined by
\begin{equation}\label{sob-inner-norm}
\langle u,v\rangle:=\langle u',v'\rangle_2\quad\mbox{and}\quad
\|u\|=\|u'\|_2,\quad u,v\in\scH.
\end{equation}
Note that $\|\cdot\|$ is by default only a semi-norm since all constant functions yield a 0 value. However, it is understood in $\sob$ that every of its functions are defined only up to a constant component.
\line
\begin{proposition}\label{prop:orthoG}
With the following subspaces of $\scH$,
\begin{align}
\scL&:=\big\{v\in\scH: v(t) \mbox{ is linear on $[t_{k-1},t_k],~\forall k\in\ZZ$}\big\},\label{scL}\\
\scC&:=\big\{w\in\scH: (w(t_k))\inZZ \mbox{ is constant}\big\},\nonumber
\end{align}
\begin{enumerate}[label=(\roman*)]
\item $\scH=\scL\overset{\scriptscriptstyle\perp}{\oplus}\scC$,\vspace{1mm}
\item $LS=P_\scL$ in $\scH$.
\end{enumerate}
\end{proposition}
\line
\begin{IEEEproof}
Let $v\in\scL$ and $w\in\scC$. For any given $k\in\ZZ$, $v'(t)$ is equal to a constant $\v_k$ on $(t_{k-1},t_k)$. Then,
$$\textstyle\int\limits_{t_{k-1}}^{t_k\!}\!v'(t)w'(t)\,\dif t=\v_k\!\int\limits_{t_{k-1}}^{t_k}\!\!w'(t)\,\dif t=\v_k(w(t_k){-}w(t_{k-1})=0.$$
By summing the above terms over $k\in\ZZ$, we obtain that $\langle v,w\rangle=\langle v',w
'\rangle_2=0$. This proves that $\scL\perp\scC$ in $\scH$.
For any $u\in\scH$, we have $u=Lu+(u{-}Lu)$ where $Lu\in\scL$ and $u{-}Lu\in\scC$ since $u(t_k)-Lu(t_k)=0$ for all $k\in\ZZ$. This shows that $\scH=\scL+\scC$, which proves (i). This simultaneously proves that $LS=P_\scL$.
\end{IEEEproof}
\ppnoi
The set $\scC$ can also be presented as the space of functions of $\scH$ that have zero crossings at $(t_k)\inZZ$ up to a constant component. Via \eqref{groch-proj}, we have thus shown that iteration \eqref{groch-alg0} is in $\sob$ equivalent to
\begin{equation}\label{grogh-sob}
x\up{n+1}= x\up{n}+\'P_\scB(w-P_\scL x\up{n})\qquad n\geq0
\end{equation}
where $w:=L\vx$. This is the POCS iteration of Table \ref{tab}\'(ii) with $\scH:=\sob$, $\scU:=\scB$ and $\scV:=\scL$.

\subsection{Impact of POCS approach}

By equivalence to \eqref{grogh-sob} in $\sob$, we know that iteration \eqref{groch-alg0} is systematically convergent in the noise-free case, up to a constant component, without any condition on the sampling separation sequence $(\Delta t_k)\inZZ$. Not only its supremum $\Delta$ is allowed to be larger than 1, but it can also be infinite. By adding the fact that the linear interpolator $L$ used in \eqref{groch-alg0} is exact with no dc offset, it can actually be proved when $\Delta$ is finite that $x\up{n}$ converges uniformly to an exact bandlimited interpolator of the points $(t_k,\x_k)\inZZ$. So perfect reconstruction is achieved not only when $\Delta<1$, but more generally whenever the sample points yield only a single bandlimited interpolator. When the reconstruction is not unique by lack of samples, $x\up{n}$ tends to the bandlimited interpolator that is closest to $x\up{0}$ with respect to the Sobolev norm defined in \eqref{sob-inner-norm}. In the presence of sampling noise and in stable conditions, the limit of $x\up{n}$ is the least-squares solution of the equation $P_\scL u=L\vx$ in $\scB$ that is closest to $x\up{0}$, again in the Sobolev sense.

\subsection{Discrete-time iteration implementation}

As implied in the introduction, it is possible to present \eqref{grogh-sob} as the POCS iteration resulting from samples of the type \eqref{samp-gen-d}. One proceeds by choosing as orthogonal basis $(v_k)\inZ$ of $\scL$ the functions of $\sob$ of derivatives $(1_{[t_{k-1},t_k)})\inZ$. One then finds that \eqref{samp-gen-d} is achieved with $\w_k:=\x_k-\x_{k-1}$. According to Sections \ref{subsec:ortho-samp} and \ref{sec:POCS}, the resulting POCS iteration for reconstruction is given by \eqref{POCS-proj} with $\scU=\scB$, $\scV=\scL$ and the vector $w$ given in \eqref{wd}. This leads to \eqref{grogh-sob} after verifying that this vector $w$ does coincide with $L\vx$ as used in \eqref{groch-projb}. From
Sections \ref{subsec:ortho-samp} and \ref{subsec:gradient}, one can then construct the  discretized version \eqref{sys} of \eqref{grogh-sob}. The problem is that the  functions  $(v_k)\inZ$ have infinite time supports regardless of their chosen dc components.

There is in fact a direct and better way to discretize \eqref{groch-alg0}, which amounts to using the non-orthogonal basis $(q_k)\inZZ:=(L\ve_k)\inZZ$ of $\scL$, where $\ve_k$ is the standard basis vector of $\RR^\ZZ$ with 1 as the $k$th coordinate and 0 elsewhere. One starts by thinking of \eqref{groch-alg0} as iteration \eqref{Landweber2} where $V$ and $V^*$ play the roles of $S$ and $P_\scB L$, respectively. Following the algebraic procedure of Section \ref{subsec:discrete}, one then finds that \eqref{groch-alg0} is equivalent to iterating the system
\begin{align*}
\vc\up{n+1}&=\vc\up{n}+  (\vx_0- S P_\scB L\' \vc\up{n})\\
x\up{n}&=x\up{0}+P_\scB L\'\vc\up{n}
\end{align*}
starting from $\vc\up{0}=0$, where $\vx_0:=\vx-S x\up{0}$. The operator $S P_\scB L$ is again a matrix.  By construction of $(q_k)\inZZ$,
$P_\scB L(\c_k)\inZZ=P_\scB \sum_{k\in\ZZ}\c_k\'q_k=\sum_{k\in\ZZ}\c_k\'P_\scB q_k$. It then follows that
\begin{equation}\nonumber
S P_\scB L=\Big[(P_\scB q_\ell)(t_k)\Big]_{(k,\ell)\in\ZZ\times\ZZ}.
\end{equation}
Using the notation $\Delta a_k:=a_k-a_{k-1}$ whether $(a_k)\inZZ$ is a sequence of scalars or functions, we show in Appendix \ref{app-discrete} that
\begin{equation}\label{Grochenig-discrete}
P_\scB q_k=\Delta\varphi_{k+1}/\Delta t_{k+1}-\Delta\varphi_k/\Delta t_k,\qquad k\in\ZZ.
\end{equation}
where $\varphi_k(t):=\varphi(t-t_k)$ and $\varphi(t)$ is defined in \eqref{phi}. For each $(k,\ell)\in\ZZ\times\ZZ$, each coefficient $(P_\scB q_\ell)(t_k)$ of the matrix $S P_\scB L$ requires the the values of $\varphi(t_j\!-t_\ell)$ for $j=k{-}1,k,k{+}1$. Like in Section \ref{subsec:mult-discr}, the values of $\varphi(t)$ can be precalculated at the discrete locations of the time-quantized grid in which the instants $(t_k)\inZ$ are acquired, and stored in a lookup table.

\subsection{Numerical experiments}

We have chosen to demonstrate the numerical behavior of Gr\'ochenig's algorithm in situations that were not considered in \cite{Grochenig92b}, i.e., in the condition of oversampling with sampling noise and under insufficient sampling.

\subsubsection{Oversampling with sampling noise}

\begin{figure}[!t]
\centering
\includegraphics[width=1\linewidth]{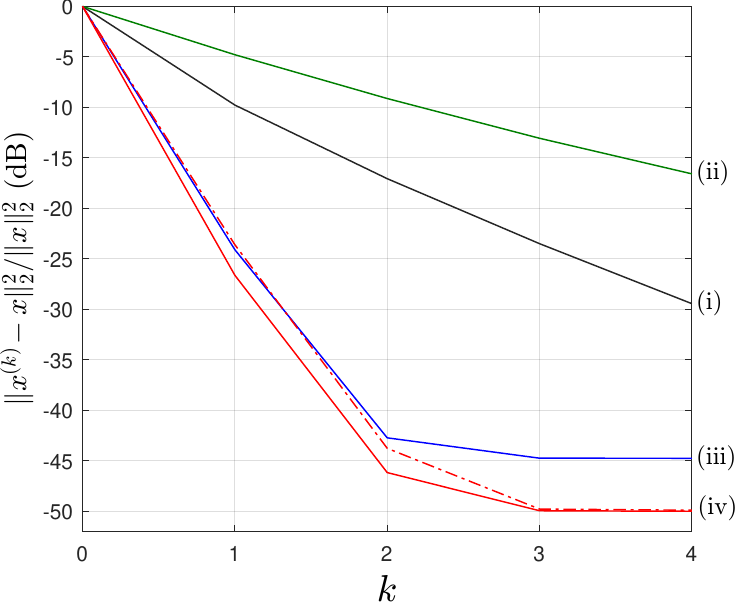}
\caption{MSE performance for several iterative algorithms (oversampling ratio $>1$), with sample values corrupted by additive noise at about $45\,\text{dB}$ below the input level. This experiment focuses on how different methods balance fast  convergence and effective noise shaping. The solid lines correspond to MSE based on the $L^2$-norm (as reported in the ordinate description), while the mixed line corresponds to MSE based on the Sobolev norm.}
  \label{fig2}
\end{figure}

We plot in Fig. \ref{fig2} the MSE performance of a number of iterative bandlimited reconstruction algorithms from nonuniform point samples of similar complexity per iteration, in an overall condition of oversampling with noise on the samples. The reconstruction  methods include the original frame algorithm introduced in the pioneering paper on nonuniform sampling \cite{Duffin52} (curve (i)), the cyclic Kaczmarz method (curve (ii)) \cite{Yeh90}, its randomized version \cite{strohmer2009randomized} (curve (iii)), and Gr\"ochenig's algorithm (curve (iv)). For each algorithm, the MSE of the $n$th iterate $x\up{n}$, reported in solid lines, is measured by averaging the relative error $\|x\up{n}\!-x\|_2^2/\|x\|_2^2$ over 100 randomly generated bandlimited inputs $x$ that are periodic of period 315 (assuming a Nyquist period 1). For each input, the sampling instants $(t_k)\inZ$ are also randomly generated such that $(\Delta t_k)\inZ$ is generated as an i.i.d sequence that is uniformly distributed in $[0,0.5]$ (in Nyquist period unit) and the sample errors are Gaussian random variables that are 45 dB's below the input-signal level. The resulting high oversampling ratio has been chosen to highlight the noise-shaping effect of the POCS algorithm. Even though our analysis of Gr\"ochenig's algorithm has been constructed in the Sobolev space $\sob$, we have maintained the $L^2$-norm in the error measurements as this is the standard reference of MSE in signal processing. We have however superimposed in mixed lines the MSE obtained by averaging $\|x\up{n}\!-x\|^2/\|x\|^2$ where $\|\cdot\|$ is the norm of $\sob$ given in \eqref{sob-inner-norm}, specifically for Gr\"ochenig's algorithm (curve (iv)). Even though the two norms are not equivalent, we observe that they yield similar results. So, while we do not provide analytical justifications for this similarity, we see that the Sobolev norm remains an adequate tool of error predictions in these experiments. Meanwhile, the plot shows the superior convergence of Gr\"ochenig's method over the other methods in the presence of noise. This is a consequence of the least-squares resolution when interpreted as a POCS iteration.

\subsubsection{Sub-Nyquist situation}\label{subsec:exp2}

\begin{figure}[!t]
  \centering
\centerline{\hbox{\scalebox{0.7}{\includegraphics{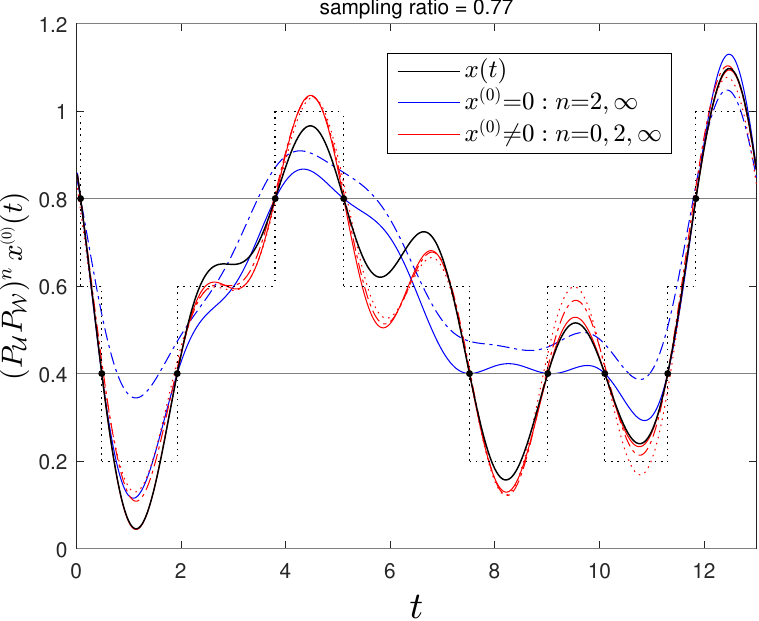}}}}
\caption{Sub-Nyquist level-crossing sampling. The true  bandlimited signal $x(t)$ (black) crosses horizontal levels  (gray). Two sets of POCS reconstructions are shown with  different initial guesses $x\up{(0)}(t)$: zero (blue) versus a  piecewise-constant approximation (red). Dotted lines:  initial guess ($n=0$); solid lines: after 2 iterations ($n=2$);  dashed lines: final solution ($n=\infty$).}  \label{fig3}
\end{figure}
We show in Fig. \ref{fig3} an example of POCS iteration limit in a case of sub-Nyquist sampling. In this experiment, the samples of the bandlimited input $x(t)$ are obtained by level-crossing sampling \cite{Marvasti01,Miskowicz2018,Rzepka18}, that is, from the crossings of $x(t)$ with fixed levels represented by horizontal grey lines. The resulting sampling ratio is $0.77$ (the time unit is the Nyquist period of $x(t)$), which prevents uniqueness of reconstruction. While the blue curves result from a zero initial estimate $x\up{0}(t)$, the red curves are obtained by choosing for $x\up{0}(t)$ the bandlimited version of the piecewise constant function shown in black dotted curve. This stair case function can be generated from the mere knowledge of the level crossings. In each of the two cases of initial estimate, the result of 3 iteration numbers is plotted, using the following line types in increasing order of iteration: dotted line, mixed line, solid line. The exact iteration numbers of the plots are indicated in the legend. To validate our convergence analysis, we have actually plotted the result of infinite iteration by matrix pseudoinversion, a technique that is made possible by the finite dimensionality of the experiment. The figure shows the good convergence of $x\up{n}(t)$ to this ideal limit. Now, the main point of this figure is to show an example of reconstruction  improvement using a heuristically designed nonzero initial estimate $x\up{0}(t)$.

\section{Conclusion}

In this paper, we have developed a whole theory of signal reconstruction when, in a Hilbert space $\scH$, data is extracted from an input $x\in\scU$ in the form $w=P_\scV x$ with some possible acquisition deviation $d$. In this case, there is a POCS-based algorithm for the reconstruction of $x$, which systematically tends to the least-squares solution of the equation $P_\scV u =w$ in $\scU$ that is closest to any chosen initial guess $x\up{0}$, provided that the least-squares solutions theoretically exist. As a result, it achieves best input approximations under noisy or insufficient sampling, while guaranteeing perfect reconstructions whenever $d=0$ and $P_\scV u=w$ has a unique solution in $\scU$. Thus, the POCS algorithm addresses simultaneously the practical concerns of non-ideal sampling and the theoretical objectives of perfect reconstruction. When the acquired data is in the form of samples $(\w_k)\inZ$, the knowledge of $w:=P_\scV x$ is inherently obtained when $\scV$ has an orthogonal basis $(v_k)\inZ$ and $\w_k=\langle x,v_k\rangle$ for all $k\in\Z$. In this case, the iterative part of the POCS algorithm has an intrinsic and rigorous discrete-time implementation, which does not require a Nyquist-rate type of sampling of the iterates. This situation of data acquisition appears in a number of emerging event-based sampling schemes based on input integration (ASDM, integrate-and-fire, multi-channel time encoding). Through a more fundamental analysis, our framework either leads to improved reconstruction methods compared to those previously proposed, or reveals convergence properties of the existing methods under noisy or insufficient sampling — properties that have remained unknown until now. We have also applied our framework to an existing algorithm for the bandlimited interpolation of nonuniform point samples, and in this process, have uncovered convergence properties of this method beyond the ideal situation of perfect reconstruction. With these applications, we have not only demonstrated the effectiveness of our framework on existing encoding schemes, but we have also illustrated its potential for generating new event-based encoders under the constraint of a robust and readily implementable signal reconstruction method.
Our approach to multi-channel time encoding fundamentally consisted in extending the principle of integrating the input over disjoint parts found in ASDM and integrate-and-fire, from one-dimensional integration to the multi-dimensional integration of functions on any measurable space. In the second application of point sampling, we have shown the effectiveness of our framework in dealing with ambient Hilbert spaces other than the traditional $L^2$-spaces.

\appendix

\subsection{Proof of Theorem \ref{theo:sep}}\label{app:sep}

Theorem \ref{theo:sep} is based on the following property of the space $\scU$ defined in \eqref{U-fib}.
\line
\begin{lemma}\label{lem2}
Let $\v\in\RR^M$ and $v\in L^2(\RR)$. If $\v\in\rmA^\perp$ or  $v\in\scB^\perp$, then $\v\otimes v\in\scU^\perp$.
\end{lemma}
\line
\begin{IEEEproof}
Let $\bu\in\scU$. Assume that $\v\in\rmA^\perp$. It follows from \eqref{mult-inprod} that
$$\langle \v\otimes v, \bu \rangle = \int_\RR\!v(t)\Big(\smallsum{\ell\in\M}\v(\ell)\'\bu(\ell, t)\Big)\,\dif t.$$
Since $\v\in\rmA^\perp$ while $\bu(\cdot,t)\in\rmA$ for all $t\in L^2(\RR)$, then
$\sum_{\ell\in\M} \v(\ell) \bu(\ell, t)= 0$ for all $t\in L^2(\RR)$. Thus, $\langle \v\otimes v, \bu \rangle=0$. Assuming that $v\in\scB^\perp$ one obtains the same result using the fact that $\bu(\ell,\cdot)\in\scB$ for all $\ell\in\M$, and the expansion
$$\langle \v\otimes v, \bu \rangle = \smallsum{\ell\in\M}\v(\ell)\Big(\int_\RR\!v(t)\'\bu(\ell, t)\,\dif t\Big).$$
This suffices to prove that $\v\otimes v \in\scU^\perp$.
\end{IEEEproof}
\ppnoi
We now proceed with the proof of Theorem \ref{theo:sep}.
Decompose $\v = P_{\rmA} \v + \w$ and $v = P_{\scB} v + w$ with $\w\in\rmA^\perp$ and $w\in\scB^\perp$. Then $\v\otimes v= P_{\rmA}\v \otimes P_{\scB}v
+ \w \otimes P_{\scB}v+P_{\rmA}\v\otimes w+ \w \otimes w$.
By Lemma \ref{lem2}, the last three terms lie in $\scU^\perp$. Since the first term $P_{\rmA}\v\otimes P_{\scB}v$ is in $\scU$, it is equal to $P_\scU(\v\otimes v)$.

\subsection{Proof of \eqref{Karen-discrete} and \eqref{Grochenig-discrete}}\label{app-discrete}

For short notation, we will write $\int_a^b f(t)\dif t=\int_a^b f$ whenever possible.
Via the Cauchy formula for repeated integration. the function $\varphi$ of \eqref{phi} has the equivalent expression
\begin{equation}\nonumber
\textstyle\varphi(t)=\int\limits_0^t\int\limits_0^s\sinc(\tau)\dif\tau\dif s=\int\limits_0^t\rmS\quad\mbox{where}\quad\rmS(s):=\int\limits_0^s\sinc.
\end{equation}

\subsubsection{Proof of  \eqref{Karen-discrete}}
~\vspace{-2mm}
\begin{equation}\label{inn-P11}
\big\langle P_\scB 1_{[a,b]},1_{[c,d]}\big\rangle=\textstyle\int\limits_0^d P_\scB 1_{[a,b]}-\int\limits_0^c P_\scB 1_{[a,b]}.
\end{equation}
\vspace{-4mm}
\begin{align*}
(P_\scB 1_{[a,b]})(t)&=\textstyle\int\limits_{-\infty}^\infty\sinc(s) 1_{[a,b]}(t{-}s)\dif s=\int\limits_{t-b}^{t-a}\sinc\\
&=\rmS(t{-}a)-\rmS(t{-}b),
\end{align*}
\vspace{-6mm}
\begin{eqnarray*}
&\textstyle\int\limits_0^\beta \rmS(t{-}\alpha)\'\dif t=\int\limits_{-\alpha}^{\beta-\alpha}\rmS=\varphi(\beta-\alpha)-\varphi(-\alpha),\\
&\textstyle\int\limits_0^\beta P_\scB 1_{[a,b]}=\varphi(\beta-a)-\varphi(-a)-\varphi(\beta-b)+\varphi(-b),
\end{eqnarray*}
One obtains \eqref{Karen-discrete} by injecting this into \eqref{inn-P11} with $\beta=d,c$.
\line

\subsubsection{Proof of \eqref{Grochenig-discrete}}

Let $\vone(t)$ be the unit step function, $r(t):=(\vone\!*\!\vone)(t)$ be the ramp function, and $r_k(t):=r(t{-}t_k)$. It can be seen that $q_k=\Delta r_{k+1}/\Delta t_{k+1}-\Delta r_k/\Delta t_k$, so that
\begin{equation}\label{PBq}
P_\scB q_k=  P_\scB\Delta r_{k+1}/\Delta t_{k+1}- P_\scB\Delta r_k/\Delta t_k,\qquad k\in\ZZ.
\end{equation}
Since $-\Delta r_k=\vone*1_{I_k}$ where $I_k:=[t_{k-1}t_k)$, then $-P_\scB\Delta r_k=(\sinc\!*\!\vone)*1_{I_k}$. We have
$$(\sinc\!*\!\vone)(t)=\textstyle\int\limits_{-\infty}^t\sinc=\int\limits_{-\infty}^0\sinc+\int\limits_0^t\sinc
=\smallfrac{1}{2}+\rmS(t),$$
\vspace{-4mm}
\begin{align*}
-(P_\scB\Delta r_k)(t)=\textstyle\int\limits_{-\infty}^\infty(\sinc\!*\!\vone)(s)1_{I_k}(t{-}s)\dif s=\!\!\int\limits_{t-t_k}^{t-t_{k-1}}\!\!(\sinc\!*\!\vone)\\
=\textstyle\smallfrac{\Delta t_k}{2}+\!\!\int\limits_{t-t_k}^{t-t_{k-1}}\!\!\!\rmS=\smallfrac{\Delta t_k}{2}+\varphi(t-t_{k-1})-\varphi(t-t_k).
\end{align*}
Thus, $P_\scB\Delta r_k/\Delta t_k=-\frac{1}{2}+\Delta\varphi_k/\Delta t_k$. Then \eqref{PBq} leads to \eqref{Grochenig-discrete}.

\bibliographystyle{ieeetr}

\bibliography{reference}{}

\end{document}